# L'IA du Quotidien peut elle être Éthique ?

Loyauté des Algorithmes d'Apprentissage Automatique

Philippe Besse[1], Céline Castets-Renard[2], Aurélien Garivier[3] & Jean-Michel Loubes[4]

**Résumé**

Associant données massives (*big data*) et algorithmes d'apprentissage automatique (*machine learning*), la puissance des outils de décision automatique suscite autant d'espoir que de craintes. De nombreux textes législatifs européens (RGPD) et français récemment promulgués tentent d'encadrer les usages de ces outils. Laissant de côté les problèmes bien identifiés de confidentialité des données et ceux d'entrave à la concurrence, nous nous focalisons sur les risques de discrimination, les problèmes de transparence et ceux de qualité des décisions algorithmiques. La mise en perspective détaillée des textes juridiques, face à la complexité et l'opacité des algorithmes d'apprentissage, révèle la nécessité d'importantes disruptions technologiques que ce soit pour détecter ou réduire le risque de discrimination ou pour répondre au droit à l'explication. La confiance des développeurs et surtout des usagers (citoyens, justiciables, clients) étant indispensable, les algorithmes exploitant des données personnelles se doivent d'être déployés dans un cadre éthique strict. En conclusion nous listons, pour répondre à cette nécessité, quelques possibilités de contrôle à développer : institutionnel, charte éthique, audit externe attaché à la délivrance d'un label.

**Mots clefs**

Intelligence artificielle, éthique, apprentissage automatique, discrimination, *discrimination impact assessment,* droit à l'explication, qualité d'une décision automatique.

---

[1]  Université de Toulouse INSA, Institut de Mathématiques UMR CNRS 5219.

[2]  Université Toulouse Capitole, Institut de Recherche en Droit Européen, International et Comparé (IRDEIC). Membre de l'Institut Universitaire de France (IUF).

[3]  Ecole Normale Supérieure de Lyon, UMPA UMR CNRS 5569.

[4]  Université de Toulouse Paul Sabatier, Institut de Mathématiques UMR CNRS 5219.

# *Can Everyday AI be Ethical?*
## *Machine Learning Algorithm Fairness*

**Abstract**

Combining big data and machine learning algorithms, the power of automatic decision tools induces as much hope as fear. Many recently enacted European legislation (GDPR) and French laws attempt to regulate the use of these tools. Leaving aside the well-identified problems of data confidentiality and impediments to competition, we focus on the risks of discrimination, the problems of transparency and the quality of algorithmic decisions. The detailed perspective of the legal texts, faced with the complexity and opacity of the learning algorithms, reveals the need for important technological disruptions for the detection or reduction of the discrimination risk, and for addressing the right to obtain an explanation of the automatic decision. Since trust of the developers and above all of the users (citizens, litigants, customers) is essential, algorithms exploiting personal data must be deployed in a strict ethical framework. In conclusion, to answer this need, we list some ways of controls to be developed: institutional control, ethical charter, external audit attached to the issue of a label.

## 1 Introduction

En 2017 la CNIL a initié un débat national sur le thème : « Ethique numérique : les algorithmes en débats » qui a donné lieu à la [publication d'un rapport](). Besse et al. (2017) y avait apporté leur [contribution]() dont cet article reprend et développe la section 4 pour continuer à faire avancer la réflexion sur le thème de la l*oyauté des décisions algorithmiques*.

Mandatée par le gouvernement, une commission présidée par Cédric Villani a elle aussi publié un rapport dans l'objectif est de « [donner du sens à l'Intelligence artificielle]() » (IA). Comme le rapport de la CNIL, le rapport de la commission Villani consacre une large place aux questions éthiques soulevées par la généralisation de l'usage, au quotidien, d'algorithmes d'Intelligence Artificielle. La France n'est évidemment pas seule à se mobiliser sur cette question et les initiatives sont nombreuses dont celle du gouvernement britannique qui a publié un [cadre éthique des données]().

Il n'est pas question d'aborder l'ensemble des algorithmes du vaste champ disciplinaire de l'IA mais de se focaliser sur ceux conduisant à des décisions impactant les personnes au quotidien : accès à la banque, l'assurance, la santé, l'emploi, les applications en matière judiciaire ou de police... Plus précisément cela concerne les algorithmes dits d'apprentissage automatique (*machine learning*) entraînés sur de vastes ensembles de données à minimiser certains critères mathématiques ou plus précisément statistique comme un taux d'erreur moyen afin d'automatiser la production de décisions.

Schématiquement les questions éthiques concernent principalement les problèmes de confidentialité des données à la base de l'apprentissage, d'entrave à la concurrence, de transparence ou *explicabilité* des décisions, de leurs risques de biais discriminatoires envers des individus ou groupes sensibles.

Avec le déploiement du RGPD (Règlement Général européen sur la Protection des Données n°2016/679/UE), la CNIL focalise son action sur son cœur de métier, c'est-à-dire plus précisément sur la protection des données personnelles en proposant aux entreprises concernées des outils de mesure à même d'évaluer les risques encourus en matière de confidentialité : le DPIA ou *[data protection impact assessment](#)*. Il appartient en effet aux entreprises d'être proactives sur ce sujet pour être en mesure de montrer, en cas de contrôle, qu'elles maîtrisent la sécurité des données personnelles dans toute la chaîne de traitement, de l'acquisition à la décision. La constatation de défaillances sera l'occasion de très lourdes sanctions financières : jusqu'à 20M€ et majorée pour une entreprise à 4 % du chiffre d'affaire annuel mondial (le plus élevé des deux chiffres devant être retenu).

De son côté, suite à l'adoption de la loi n° 1321-2016 pour une République numérique, qui tient notamment compte de quelques dispositions du RGPD, l'INRIA a proposé un projet de plateforme collaborative ([TransAlgo](#)) qui permettrait d'archiver des outils automatiques produits par cinq groupes de travail :
1. Moteurs de classement de l'information et systèmes de recommandation ;
2. Apprentissage : robustesse aux biais des données et des algorithmes, reproductibilité, explication et intelligibilité ;
3. Protection des données et contrôle d'usage des données ;
4. Métrologie des réseaux de communication ;
5. Influence, désinformation, impersonification (photos, voix, agent conversationnel), *nudging, fact-checking*.

Nous proposons dans cet article des éléments de réflexion et outils pour faire avancer le point 2 sur les risques de discrimination, ainsi que l'explicabilité, la répétabilité ou la qualité des décisions algorithmiques ou automatiques.

- *Discrimination* : La loi protège les individus contre des pratiques discriminatoires, mais comment peut-elle être opposée à des algorithmes ? Elle n'évoque pas la discrimination de groupe mais le rapport Villani appelle de ses vœux, dans la section 5, la création d'un DIA (*discrimination impact assessment)* sans pour autant faire référence à une littérature américaine déjà abondante sur le sujet. Quels sont les outils disponibles à ce propos ?
- *Explicabilité* : L'analyse fine des textes juridiques montre que les obligations légales sont relativement peu contraignantes en matière de transparence des algorithmes. Néanmoins, l'acceptation de l'IA et de décisions automatiques impactant des personnes, requiert impérativement des éléments de transparence ; c'est dans ce cas le *droit à l'explication* d'une décision algorithmique. Quels peuvent en être les termes ?
- *Qualité:* la loi française, comme le RGPD, n'évoquent à aucun moment des notions de qualité ou risque d'erreur d'une décision automatique. Comme pour les sondages, il serait pertinent que la loi oblige à informer l'usager des risques associés à l'exécution d'un algorithme d'apprentissage automatique. Quel en est le contexte ?

Comme le rappelle le rapport Villani, les thèmes de l'éthique ont investi l'espace entre ce que permettent les nouvelles technologies issues de l'IA et ce qui est permis par la loi ; il insiste en notant que le « temps du droit est bien plus long que celui du code ». Aussi, en l'absence de textes législatifs plus précis, alors que la notion de loyauté d'une plateforme est présente dans la loi pour une République numérique, les principes de loyauté des algorithmes deviennent des questions éthiques et juridiques, pas simplement par altruisme de la part des entreprises commerciales, mais pour le développement d'une confiance indispensable du grand public envers le déploiement de ces technologies. Appréhender les problèmes soulevés par la généralisation du compteur Linky, ceux liés à la mise en place de ParcoursSup ou encore les déboire en bourse de Facebook à la suite de l'affaire *Cambridge Analytica* en sont de bons exemples.

Il importe, en premier lieu, de pouvoir mieux définir comment des notions d'éthique peuvent se traduire en termes techniques.

La section 2 suivante décrit plus précisément les algorithmes d'apprentissage statistique, branche de l'IA, concernés par cet article. La section 3 décrit le contexte juridique et les moyens disponibles pour un individu ou un groupe de se protéger en basant la définition d'une mesure d'impact disproportionné (*disparate impact*) de discrimination sur celle de la littérature et ses récents développements, notamment pour corriger un biais d'apprentissage. La section 4 aborde le droit à l'explication au regard des capacités techniques des modèles statistiques et algorithmes d'apprentissage très généralement utilisés. La section 5 rappelle comment sont estimés et minimisés des risques d'erreur ; les conséquences d'une erreur de 30 % ne sont pas les mêmes lorsqu'il s'agit de l'évaluation du risque de récidive d'un détenu ou de celle de l'intérêt d'un client de *Netflix* pour un film. Enfin, après avoir tenté de résumer une situation pour le moins complexe, nous concluons en évoquant les quelques possibilités de contrôle institutionnel, d'autocontrôle (charte éthique) ou encore de contrôle externe (audit) attaché à la délivrance d'un label.

## 2 Quelle IA ? Quels algorithmes ?

L'IA couvre un vaste champ disciplinaire et concernent donc de nombreux types d'algorithmes. Nous nous intéresserons plus particulièrement à ceux couramment mis en œuvre dans notre quotidien et conduisant à des décisions à fort impact et construites à partir de données personnelles. Il s'agit donc d'algorithmes dits d'*apprentissage automatique* (*machine learning*) d'une décision à partir d'un historique de situations connues ou observées sur un plus ou moins grand échantillon dit d'apprentissage ou d'entraînement. Parmi ceux-ci, laissons de côté les algorithmes par renforcement (cf. AlphaGo) ou séquentiels dont les applications concernent plus particulièrement le commerce en ligne (algorithmes de bandits) aux conséquences moins critiques. Nous allons donc nous focaliser sur un sous-ensemble, nommé *apprentissage statistique* des algorithmes d'apprentissage automatique.

## 2.1 Exemples d'utilisation

Le choix d'un traitement médical, d'une action commerciale, d'une action de maintenance préventive, d'accorder ou non un crédit, de surveiller plus particulièrement un individu... toutes les décisions qui en découlent sont la conséquence d'une prévision. La prévision du risque ou de la probabilité de diagnostic d'une maladie, le risque de rupture d'un contrat par un client qui est le score d'attrition (*churn*), le risque de défaillance d'un système mécanique, de défaut de paiement d'un client ou encore de radicalisation d'un individu... Les exemples sont très nombreux et envahissent notre quotidien. Ces prévisions de risques ou scores, par exemple de crédit, sont produits par des algorithmes d'*apprentissage statistique*, après entraînement sur une base de données.

## 2.2 Algorithmes d'*apprentissage statistique*

Dans les années 30 et notamment à la suite des travaux de Ronald Fisher, la Statistique a été développée avec une finalité *principalement explicative* pour un objectif d'aide à la décision. Par exemple : tester l'efficacité d'une molécule et donc d'un médicament, comparer le rendement de semences ou optimiser le choix d'un engrais, montrer l'influence d'un facteur (consommation de tabac, de sucre) sur des objectifs de santé publique. La prise de décision est alors la conséquence d'un *test statistique* permettant de contrôler le *risque d'erreur* encouru. Mais il se trouve que les mêmes modèles statistiques peuvent aussi être utilisés avec une *finalité* seulement *prédictive* : prévoir la concentration en ozone du lendemain, le risque de défaut de paiement d'une entreprise… De plus, ces modèles statistiques peuvent être suffisamment simples, (typiquement linéaires) pour être facilement *interprétables*.

Néanmoins, certaines situations, certains phénomènes, nécessitent des modèles, ou plus généralement des algorithmes, plus complexes pour être correctement approchés afin de conduire à des prévisions suffisamment fiables. Toutes les disciplines scientifiques : Statistique et aussi Mathématiques, Informatique, ont été mise à contribution depuis la fin des années 90 pour développer une farandole d'algorithmes avec une *finalité* essentiellement *prédictive* : arbres binaires de décision, *k* plus proches voisins, machines à vecteurs supports, réseaux de neurones puis apprentissage profond (*deep learning*), forêts aléatoires, *gradient boosting machine*… en sont quelques exemples. Il n'est plus question de tester l'influence d'un facteur ou l'efficacité d'un traitement ; seule compte la *qualité de la prévision*. La littérature est très vaste à ce sujet, consulter par exemple James et al. (2017) ou les vignettes pédagogiques du site [wikistat.fr](wikistat.fr).

## 2.3 Principe de l'apprentissage statistique

Le principe des algorithmes d'apprentissage statistique repose sur le fait de pouvoir élaborer, à partir d'un ensemble d'exemples appelé échantillon d'apprentissage, une règle de décision qui va s'appliquer à tous les nouveaux cas rencontrés. A partir d'un grand nombre de données recueillies, contenant principalement des décisions qui ont déjà été prises et les variables qui expliquent ses décisions, les principes mathématiques permettent non seulement de comprendre comment ces décisions ont été prises mais également de dégager les règles qui président à ces choix.

La découverte de ces règles consiste concrètement à trouver des tendances (*patterns* ou *features*) dans les observations. Pour cet apprentissage il faut trouver, donc détecter dans les données, des comportements caractéristiques qui permettent de segmenter les individus en des groupes homogènes. Ainsi donc, en fonction de nos caractéristiques, notre profil type peut être défini par rapport aux autres individus déjà analysés et l'algorithme émettra une règle fixe en fonction de notre groupe d'appartenance ou par rapport à des similarités (ressemblances) vis-à-vis des individus déjà étudiés. Le processus de découverte des comportements type est automatisé, sans contrôle *a posteriori*. Or c'est à partir de ces comportements types que se créent les modèles, se prennent les décisions et que sont prédits les événements à venir.

Plus précisément, les paramètres de modèles sont estimés ou des algorithmes sont *entraînés* sur des ensembles de données d'apprentissage (*training data sets*) et optimisés de sorte qu'ils minimisent une erreur de prévision ou erreur de généralisation. Cette erreur est très généralement estimée par le calcul d'une *moyenne statistique* c'est-à-dire par la moyenne, ou le taux moyen d'erreur, commis sur un *échantillon test* indépendant de l'échantillon d'apprentissage. Un algorithme d'apprentissage statistique s'adapte au mieux à des données historiques afin d'identifier les spécificités de données actuelles ou en cours d'acquisition ; il en déduit la prévision la plus adaptée sans possibilité de créativité : schématiquement, chercher la situation passée la plus ressemblante à la situation actuelle pour en déduire la prévision la plus fidèle.

## 2.4 Risques de l'apprentissage statistique

Il est important de noter que l'échantillon d'apprentissage doit être d'autant plus volumineux que l'algorithme est complexe au sens du nombre de paramètres à estimer et participant à sa définition. Corrélativement, l'accroissement spectaculaire des capacités de calcul et d'archivage associé à l'explosion des volumes de données disponibles ont permis des avancées très significatives, dont celle de l'apprentissage profond (*deep learning*) depuis 2012, dans la qualité des algorithmes d'apprentissage et donc des décisions qui en découlent. En ce sens, le succès actuel et le battage médiatique de l'apprentissage statistique et donc plus généralement de l'IA sont une conséquence directe de la *datafication* de notre quotidien qui vise à l'enregistrement et l'exploitation systématique de tous nos messages, recherches, achats, déplacements…

Ceci éclaire l'avènement du paradigme de l'ère du *Big Data*. Dans un raisonnement traditionnel cartésien, une théorie permet d'élaborer un modèle fruit d'une réflexion humaine. Puis ce modèle est confronté à la réalité au travers de données recueillies au cours d'expériences prévues pour confronter les données au modèle. Ainsi la théorie peut être clairement réfutée ou acceptée sur la base de faits. Le modèle peut alors être analysé d'un point de vue éthique ou moral, discuté même. Mais en apprentissage la création du modèle provient de l'étude des données, sans analyse *a posteriori*. On comprend dès lors qu'à partir du moment où nous décidons de confier à l'algorithme un pouvoir décisionnel, il peut façonner la réalité pour être conforme à son modèle. Il fige pour ainsi dire la réalité sur la base de ce qu'il en a vu à travers le prisme de l'échantillon fourni à l'apprentissage, puis il reproduit le modèle à l'infini. Naturellement, le modèle n'évolue plus et vient ajuster la réalité à sa propre prédic-tion. Le comportement étant appris, la règle de prédiction peut alors clairement être exprimée : terminée la place du hasard ou celle de la créativité, place à la répétabilité.

Souvent la confrontation des idées permet à chacun de préciser sa propre Vérité en prenant conscience de ses erreurs même si nous décidons de faire sciemment un mauvais choix. L'IA est autrement catégorique : la matrice algorithmique a pour but d'optimiser les décisions « justement ou froidement ». Naturellement la morale ou l'équité de ce jugement n'est pas prédéfinie mais dépend d'une part, de la manière avec laquelle sont apprises les règles (le critère objectif qui a été choisi) et d'autre part, de la manière avec laquelle a été constitué l'échantillon d'apprentissage. Le choix des règles mathématiques permettant de créer le modèle est primordial.

Se pose alors une question délicate : Comment définir ou par quelles caractéristiques « mesurables » traduire des notions de loyauté, confiance, responsabilité (*fairness*, *trustworthiness*, *accountability*), appliquées à de telles décisions algorithmiques lorsqu'elles sont la conséquence ou le résultat d'une prévision ?

La réponse se décline en trois points.

- elle doit éviter tout biais *discriminatoire* vis-à-vis de minorités et groupes sensibles protégés par la loi.
- Même statistique ou probabiliste, cette décision doit pouvoir être *attribuée* à un humain qui en assume la responsabilité. Il doit pouvoir en *rendre compte* et donc, pouvoir l'*expliquer* de façon compréhensible (*e.g.* médecin à son patient).
- Elle doit être la plus *juste* au sens de l'intérêt de la personne concernée et / ou globalement de la communauté, donc issue d'une *meilleure prévision*.

Reprenons et détaillons ces trois points.

## 3 Biais et discrimination

La partie 5 du rapport Villani consacrée aux questions éthiques laisse une large place aux risques annoncés de pratiques discriminatoires des algorithmes reproduisant, voire renforçant, les biais de société. Cette section va s'attacher à rappeler les notions de discrimination, individuelle et collectives dans le but de poser les bases successives de :

- la mesure de biais discriminatoire afin
- de construire les outils de leur détection,
- et même de leur possible correction.

### 3.1 Cadre juridique

Selon l'article 225-1 du code pénal : « Constitue une discrimination toute distinction opérée entre les personnes physiques sur le fondement de leur origine, de leur sexe, de leur situation de famille, de leur grossesse, de leur apparence physique, de la particulière vulnérabilité résultant de leur situation économique, apparente ou connue de son auteur, de leur patronyme, de leur lieu de résidence, de leur état de santé, de leur perte d'autonomie, de leur handicap, de leurs caractéristiques génétiques, de leurs mœurs, de leur orientation sexuelle, de leur identité de genre, de leur âge, de leurs opinions politiques, de leurs activités syndicales, de leur capacité à s'exprimer dans une langue autre que le français, de leur appartenance ou de leur non-appartenance, vraie ou supposée, à une ethnie, une Nation, une prétendue race ou une religion déterminée. »

L'article 225-2 ajoute que : « La discrimination définie aux articles 225-1 à 225-1-2, commise à l'égard d'une personne physique ou morale, est punie de trois ans d'emprisonnement et de 45 000 euros d'amende lorsqu'elle consiste :

1. à refuser la fourniture d'un bien ou d'un service ;
2. à entraver l'exercice normal d'une activité économique quelconque ;
3. à refuser d'embaucher, à sanctionner ou à licencier une personne ».

La loi française ne concerne qu'une *approche individuelle* de la notion du risque de discrimination. Le rapport Villani demande à prendre en compte la discrimination envers un *groupe* et insiste sur la nécessité de définir un outil d'évaluation. Il évoque le *Discrimination Impact Assessment* (DIA) en complément du *Data Protection Impact Assessment* (DPIA) prévu par le RGPD et qui protège les données personnelles des individus et non des groupes. Ce n'est pas du tout évoqué dans le rapport Villani mais il existe une littérature abondante sur ce sujet sous l'appellation de *disparate impact* (impact disproportionné) depuis les années soixante-dix aux USA.

De son côté, le règlement européen encadre strictement la collecte de données personnelles sensibles (orientation religieuse, politique, sexuelle, origine ethnique, …) et interdit aux responsables de décisions algorithmiques de les prendre en compte dans les traitements automatisés (art. 22§4), sous réserve du consentement explicite de la personne ou d'un intérêt public substantiel. Par opposition à discriminatoire, une décision est dite loyale si elle ne se base pas sur l'appartenance d'une personne à une minorité protégée ou la connaissance explicite ou implicite d'une donnée personnelle sensible.

Ce point est sans doute le plus difficile à clarifier. En effet, il ne suffit pas que la variable « sensible » soit inconnue ou supprimée des données d'apprentissage pour que la décision soit sans biais vis-à-vis de ses modalités. L'information sensible peut être contenue implicitement, même sans intention de la rechercher, dans les informations non sensibles et ainsi participer au biais de la décision. Des habitudes de consommation, des avis sur les réseaux sociaux, des données de géolocalisation… renseignent sur les orientations de la personne et peuvent permettre de reconstituer implicitement des données sensibles.

Les questions posées et difficultés rencontrées lors de la construction d'algorithmes avec un objectif de loyauté sont directement liées aux conditions d'apprentissage des décisions. En effet, comme il a été vu dans la section 2 précédente, l'apprentissage est le reflet de la base des données d'entraînement. En conséquence :
- si les données sont elles-mêmes biaisées, pas représentatives de la population,
- si un biais structurel est de toute façon présent dans la population,

alors ceci est la source d'une *rupture d'équité*. La décision reproduit voire même peut renforcer le biais, donc la discrimination. Plus dangereux encore, la décision devient une prévision auto-révélatrice. Une estimation (trop) élevée d'un risque de crédit génère un taux, donc des remboursements, plus élevés qui renforcent le risque de défaut de paiement. Un risque trop élevé de récidive retarde une mise en liberté, accroît la désocialisation et renforce *in fine* le risque de récidive. Cathy O'Neil (2016) développe en détail la perversité des effets de bord de ce type d'outils. Il faut noter que les algorithmes de classification ont pour objectif de séparer la population en sous-groupes. Ainsi si cette séparation pré-existe dans les données, l'algorithme apprendra et amplifiera cette dissimilitude et peut dès lors introduire une discrimination de traitement dans les données.

La figure 1 illustre un exemple sévère de biais des bases de données pangénomiques *(genome-wide association study, GWA study,* ou *GWAS)*. Ces bases archivent les analyses des variations génétiques (*singular nucleotid polymorphisms* ou SNPs) chez de nombreux individus, afin d'étudier leurs corrélations avec des traits phénotypiques, par exemple des maladies. Elles sont la première brique pour la recherche de thérapies personnalisées.

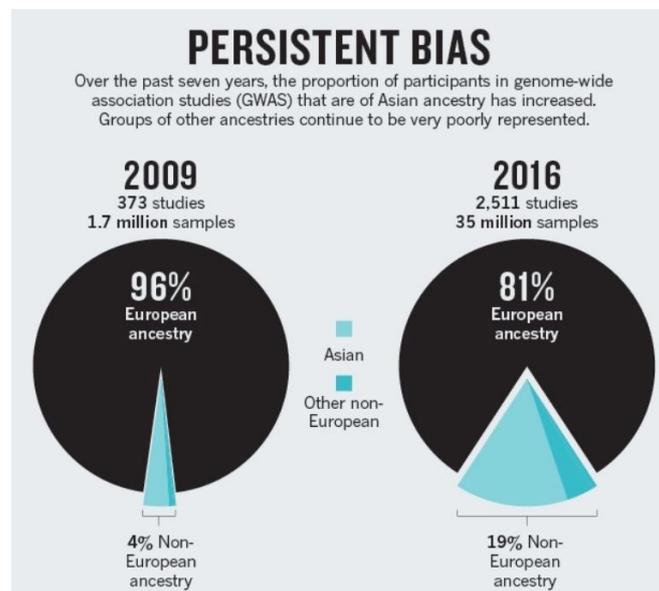

*Figure 1 : Panpejoy et Fullerton (2016) : Biais de composition des échantillons des études d'associations pangénomiques (GWAS).*

D'un point de vue ethnique, le problème est que la grande majorité des bases GWAS ont été constituées sur des populations d'ascendance blanche/européenne (cf. Figure 1, Popejoy et Fullerton, 2016). Les facteurs de risque estimés par des modèles statistiques classiques (régression logistique) ou par des algorithmes d'apprentissage automatique seront donc très probablement beaucoup moins précis pour un patient d'ascendance africaine ou asiatique. Autrement dit, ceux-ci n'ont actuellement rien à attendre d'une médecine personnalisée basée sur des considérations génomiques.

### 3.2 Discrimination individuelle

Apporter la preuve d'une discrimination individuelle est particulièrement difficile pour la personne concernée, à moins d'accepter l'utilisation de procédés probatoires en principe interdits car considérés comme déloyaux. Il s'agit de dispositifs de type « *testing* » : *test de situation* ou *test de discrimination*, qui sont à la fois des moyens d'investigation et une forme d'expérimentation sociale en situation réelle, destinés à déceler une situation de discrimination. Cette dernière peut notamment porter sur des données sensibles comme l'origine ethnique, le handicap, le sexe, l'orientation sexuelle, la religion, l'adhésion syndicale. Ce type de test ne respecte pas le principe de la loyauté de la preuve, mais il est le moyen le plus efficace, voire souvent le seul, de prouver la discrimination. Dans le cas le plus simple, il s'agit de comparer le comportement d'un tiers envers deux personnes ayant exactement le même profil pour toutes les caractéristiques pertinentes, à l'exception de celle que l'on soupçonne de donner lieu à discrimination. Naturellement, lorsque la discrimination ne repose pas sur une seule ou même plusieurs données sensibles, mais est le résultat de croisement de données permettant indirectement la discrimination, il faut pouvoir appréhender les résultats avec toute leur complexité.

Cette méthode, utilisée par les associations comme SOS racisme, est reconnue par les juridictions françaises, dans la mesure où, bien que considérée comme une pratique déloyale, elle ne peut être écartée comme moyen de recherche de la preuve depuis un arrêt de la Cour de Cassation rendu en juin 2002 dans l'affaire du Pym's de Tours. La solution a été par la suite consacrée à l'article 225-3-1 du code pénal selon lequel : « Les délits prévus par la présente section sont constitués même s'ils sont commis à l'encontre d'une ou plusieurs personnes ayant sollicité l'un des biens, actes, services ou contrats mentionnés à l'article 225-2 dans le but de démontrer l'existence du comportement discriminatoire, dès lors que la preuve de ce comportement est établie ».

Le principe est donc simple. Dans le cas par exemple d'une procédure d'embauche, il suffit d'adresser deux CV identiques à des moments différents pour répondre à des offres d'emplois. Les CV qui ne diffèrent alors que par le nom, révélateur de l'origine des candidats potentiels, ou le genre, l'âge. Pour être valide, le CV et la candidature doivent être sincères, seul le CV concurrent est modifié en conséquence. Plusieurs organismes ou laboratoires pratiquent ces opérations : citons l'Observatoire des Discriminations, le laboratoire TEPP de l'Université Marne la Vallée (L'Horty et al. 2017) ou encore la DARES (Direction de l'Animation, des Études, de la Recherche et des Statistiques) du Ministère du travail en lien avec l'association ISM Corum.... Certaines entreprises demandent d'ailleurs à ISM Corum de tester leur mode de recrutement.

Dans un autre ordre d'idée, Galhotra et al. (2017) définissent le biais d'une décision en incluant une notion de causalité. Le logiciel afférent (Themis) opère un test automatique de la possible discrimination opérée par des logiciels. Cette dernière idée peut être reprise ou plutôt simplifiée pour être appliquée aux décisions d'un algorithme pour en évaluer les risques de discriminations individuelles. Il suffit de reconsidérer un échantillon test en échangeant les deux modalités de la variable sensible ; par exemple le genre ou l'origine ethnique. L'algorithme est alors à nouveau appliqué à cet échantillon test modifié, afin d'identifier les individus pour lesquels la décision à changer, par exemple devenu plus favorable, à l'issue du seul changement de genre, d'âge ou d'origine ethnique. Même si le nombre trouvé est faible, pas statistiquement significatif, ces personnes sont clairement discriminées ; l'algorithme est source d'une sur-discrimination déloyale par rapport à des biais possibles sur l'échantillon. Ils sont potentiellement des risques de conflits juridiques coûteux pour l'entreprise.

Attention, la variable sensible doit être connue. Ce n'est évidemment pas toujours le cas et c'est un problème car supprimer la variable sensible d'un modèle n'évite pas nécessairement une décision discriminatoire mais empêche d'en identifier simplement le biais. L'autre difficulté est qu'il s'agit en plus d'apporter la preuve de *l'intention discriminatoire*. Or, s'agissant d'une discrimination par un traitement algorithmique, la discrimination n'est pas forcément le fruit d'une intentionnalité.

### 3.3 Discrimination de groupe

Le rapport Villani recommande la création ou définition d'une mesure de discrimination définie au niveau d'un groupe : *discrimination impact assessment* (DIA) et pas seulement une définition au sens individuel légal.

La première difficulté repose dans le choix d'une mesure de discrimination, alors que la littérature propose beaucoup de façons de mesurer le biais d'une décision (positive ou négative), vis-à-vis de personnes appartenant ou non à un groupe, généralement une minorité protégée par la loi. Un type de mesure individuelle s'intéresse au voisinage au sens des $k$ plus proches voisins d'un individu afin de détecter une situation atypique. Néanmoins, cet individu peut être entouré de ceux appartenant au même groupe protégé et tous ne bénéficiant pas à tort d'une décision positive. Il est plus informatif de considérer une mesure collective ou statistique de la discrimination basée sur une table de contingence (tableau 1) croisant deux variables : la *variable sensible* d'appartenance à un groupe protégé (Oui ou Non) par la loi, et l'obtention d'une décision Positive (crédit, emploi, bourse…) ou Négative.

*Tableau 1. Table de contingence entre appartenance au groupe et nature de la décision. Proportions associées : p1=a/n1, p2=c/n2, p=m1/n*

| Groupe Protégé | Décision | | |
|---|---|---|---|
| | Positive | Négative | marge |
| Oui | *a* | *b* | *n1=a+b* |
| Non | *c* | *d* | *n2=c+d* |
| marge | *m1* | *m2* | *n=n1+n2* |

Des mesures simples de discrimination sont définies à partir de cette table (Pedreschi et al. 2012) :

- Différence de risque: DR=p1-p2,
- Risque relatif: RR=p1/p2,
- Chance relative : CR=(1-p1)/(1-p2),
- Rapport de cote (*odds ratio*): RR/CR.

Mais beaucoup d'autres mesures sont proposées. Consulter par exemple Žliobaité (2015) : différences de moyennes, de coefficients de régression, tests de rangs, information mutuelle, comparaison de prévisions. Le problème est qu'il y a finalement beaucoup (trop) de définitions techniques ou statistiques possibles de la discrimination mais sans aucune base légale connue à ce jour pour en justifier le choix.

### 3.4 Impact disproportionné

Nous choisissons par la suite de nous intéresser plus particulièrement à la définition usuelle largement répandue de *disparate impact*. Celui-ci est défini comme le rapport de deux probabilités :

$$DI = \frac{P(Y=1|S=0)}{P(Y=1|S=1)}.$$

C'est le rapport de la probabilité que la décision soit positive (*Y=1*) sachant que le groupe est protégé (*S=0*) sur la probabilité que la décision soit positive (*Y=1*) sachant que le groupe n'est pas celui protégé (S=1). Il est estimé par le risque relatif (RR) défini à l'aide de la table de contingence (tableau 1 ci-dessus).

Feldman et al. (2015) fournissent quelques éléments historiques[5] de la première utilisation de ce critère par la justice de l'état de Californie et qui date de 1971. Utilisation qui a motivé de très nombreux articles dans les revues juridiques notamment à propos du mode opératoire qui consiste à comparer la valeur obtenue à un seuil fixé empiriquement à 0.8. En deçà de 0.8, l'impact est jugé suffisamment disproportionné pour être révélateur d'une discrimination. Néanmoins, si l'entreprise apporte la preuve que les choix de recrutement sont basés sur des critères nécessaires aux intérêts économiques de l'entreprise, la discrimination n'est pas attestée.

En résumé, une évaluation du DI permet, aux USA, de révéler des situations trop disproportionnées au détriment d'un groupe sensible ou protégé. Cela ouvre la possibilité de détecter voire condamner une situation de discrimination de groupe implicite ce qui est loin d'être le cas en France, ni même dans la communauté européenne, dont la législation ne reconnaît que des cas de discrimination individuelles.

L'estimation du DI soulève une autre question d'ordre statistique comme soulignée par Peresie (2009). Faut-il comparer par un test statistique l'égalité des termes du DI afin d'introduire une part d'incertitude ou simplement comparer le DI à la valeur seuil de 0.8 ? Ces deux stratégies pouvant conduire à des résultats contradictoires. De plus, le test d'égalité est basé sur une hypothèse de normalité peu judicieuse. Pour éviter ces difficultés, Besse et al. (2018) proposent une estimation du DI par intervalle de confiance incluant un contrôle statistique du risque d'erreur mais sans faire appel à une hypothèse de normalité. La distribution asymptotique exacte est obtenue en application du théorème de la limite centrale et de la linéarisation des critères de loyauté.

Malheureusement l'évaluation ou la caractérisation d'une discrimination envers un groupe ne peut se limiter à la seule évaluation du DI. Ainsi, les algorithmes de reconnaissance faciales sont régulièrement accusés de racisme mais sur le fondement d'un autre critère, celui de l'erreur de reconnaissance. Vraisemblablement à cause de bases d'apprentissage elle-même biaisées c'est-à-dire dans lesquelles certaines catégories, notamment les femmes d'ascendance africaine, sont largement sous-représentées. En conséquence, les taux d'erreur sont de l'ordre de 30% au lieu de 1% pour un homme d'ascendance européenne.

---

[5] Consulter également la page afférente du site Wikipedia en anglais.

De plus même si le DI est limité et les taux d'erreur identiques pour les catégories de la variable sensible, une autre source de discrimination peut se dissimuler dans les dissymétries des matrices de confusion d'un algorithme ou prédicateur. C'est le point de vue de l'outil utilisé par Google et rendu accessible sur la plateforme *What-If Tool*. La mesure de discrimination adoptée est alors *l'égalité d'opportunité* d'un algorithme d'apprentissage décrite par Hardt et al. (2016). Besse et al. (2018) proposent également une estimation par intervalle de confiance de cette mesure de discrimination nommée dans d'autres références : *égalité des précisions conditionnelles*.

### 3.5 Exemple : le risque de récidive *COMPAS*

Cette approche est très bien illustrée par la controverse entre le site *ProPublica* (prix Pulitzer 2011) et l'ex-société *Northpointe*, maintenant *equivant*. Cette société, dans une démarche de « justice prédictive » commercialise l'application *COMPAS* (*Correctional Offender Management Profile for Alternative Sanction*) qui produit un score ou risque de récidive pour les détenus ou accusés lors d'un procès. *ProPublica* accuse ce score d'être biaisé et donc raciste. Cette controverse a suscité de très nombreux articles venant renforcer une bibliographie déjà présente sur le sujet depuis une vingtaine d'années. Ces études mettent en évidence des contradictions rédhibitoires entre les critères proposés. Résumons la controverse.

Le score est estimé sur la base d'un questionnaire détaillé et à partir d'un modèle de durée de vie (modèle de Cox). La qualité de ce score est optimisée, mesurée, par le coefficient AUC (aire sous la courbe ROC) approximativement autour de 0.7, valeur plutôt faible correspondant aux taux d'erreurs élevés observés, de 30 à 40 %. La société *Northpointe* défend la loyauté de ce score en assurant que

- les distributions de ses valeurs (donc les taux de sélection) sont analogues selon l'origine (afro-américaine *vs.* caucasienne) des accusés ; le DI n'est pas significatif.
- le taux d'erreur sur la prévision d'une récidive (matrice de confusion) qui en découle est analogue selon l'origine, autour de 30, 40% ; l'argumentaire retenu pour la reconnaissance faciale n'est pas recevable.

De leur côté Angwin et al. (2016) du site *ProPublica* dénoncent un biais du score *COMPAS* en étudiant une cohorte de détenus libérés pour lesquels sont connus le score de récidive selon *COMPAS*, ainsi que l'observation, ou non, d'une arrestation sur une période de deux ans. Ils montrent alors que le *taux de faux positifs* : score *COMPAS* élevé mais sans récidive observée, est *beaucoup plus importants* pour les libérés d'origine afro-américaine que pour ceux d'origine caucasienne. Comme ce score *COMPAS* est utilisé pour éventuellement des remises en liberté conditionnelles, un détenu d'origine africaine a plus de malchance de rester plus longtemps emprisonné avec le risque de renforcer sa désocialisation donc finalement son risque de récidive.

Pour expliquer l'impasse de cette controverse, Chouldechova (2017) montre que, sous les contraintes de « loyauté » contrôlées par *Northpointe* et sachant que le taux de récidive des afro-américains est effectivement plus élevé, alors, nécessairement, les taux de faux positifs /

négatifs ne peuvent être que déséquilibrés au détriment des afro-américains et c'est d'autant plus manifeste que le taux d'erreur (40%) est élevé.

La question qui se pose ou qui aurait dû être posée en préalable concerne la qualité de cette prévision. Sous l'apparente objectivité d'un algorithme se cache un taux d'erreur qui discrédite largement le produit *COMPAS*. Dressel et Farid (2018) ont par ailleurs montré qu'un ensemble de personnes sans expertise judiciaire et interrogées sur le web aboutissent à des prévisions aussi (peu) fiables, de même qu'un simple modèle linéaire impliquant seulement deux variables.

### 3.6 Réparation des données pour favoriser des algorithmes loyaux.

Pour détecter la non loyauté des algorithmes, nous avons vu qu'il est possible de calculer de nombreux critères, chacun mettant en évidence un type de différence de traitement entre divers sous-groupes de la population. Il peut s'agir d'un déséquilibre entre la proportion de bonnes prédictions en fonction de chaque sous groupe ou d'une différence de répartition d'erreur ou d'autres critères qui témoignent d'une relation de dépendance entre la décision apprise et la variable qui caractérise la variable sensible et qui divise la population en deux sous-groupes. Ainsi la notion de loyauté totale doit être caractérisée par l'indépendance entre ces deux lois de probabilités (del Barrio et al. 2018b). Plus la liaison est importante, plus l'effet discriminant sera marqué. Ce formalisme a amené différents auteurs à proposer plusieurs manières de remédier à cette rupture d'équité soit en changeant la règle de décision, soit en changeant l'échantillon d'apprentissage. Modifier la règle revient à imposer à l'algorithme de ne pas sur apprendre cette liaison en imposant un terme favorisant l'absence de lien entre la prédiction et la variable dite sensible (Zafar et al. 2017). Modifier l'échantillon revient à favoriser l'indépendance entre les données et la variable sensible afin de garantir que tout algorithme utilisant ces données comme base d'apprentissage ne puisse pas reproduire un biais par rapport à la variable sensible. Pour cela, il est nécessaire de modifier les lois conditionnelles par rapport à la variable sensible et de les rendre les plus similaires possible sans perdre trop d'information qui pourrait nuire au pouvoir prédictif du modèle. Cette solution décrite dans Feldman et al. (2015), a été étudiée par del Barrio et al. (2018a). Obtenir la non-discrimination représente néanmoins un prix à payer : construire une règle moins prédictive par rapport à l'échantillon d'apprentissage. Le statisticien doit donc contrôler à la fois l'erreur commise par la règle de prédiction ainsi que la non discrimination souhaitée.

## 4 Explicabilité d'une décision

### 4.1 Lois et enjeux du droit à l'explication

Le rapport Villani appelle à « ouvrir les boîtes noires » de l'IA car une grande partie des questions éthiques soulevées tiennent de l'opacité de ces technologies. Compte tenu de leur place grandissante, pour ne pas dire envahissante, le rapport considère qu'il s'agit d'un enjeu démocratique.

L'article 10 de la loi n° 78-17 relative à l'informatique, aux fichiers et aux libertés du 6 janvier 1978 prévoyait à l'origine que « Aucune décision produisant des effets juridiques à l'égard d'une personne ne peut être prise sur le seul fondement d'un traitement automatisé de données destiné à définir le profil de l'intéressé ou à évaluer certains aspects de sa personnalité ». Autrement dit, une évaluation automatisée des caractéristiques d'une personne conduisant à une prise de décision ne peut être réalisée sur la seule base du traitement automatisé.

Cela suppose donc que d'autres critères soient pris en compte ou encore que d'autres moyens soient utilisés. En particulier, les personnes concernées par la décision peuvent attendre que l'évaluation puisse être vérifiée par une intervention humaine. Si ce principe qui tend à contrôler les effets négatifs du profilage est consacré depuis longtemps, son énoncé n'a pu empêcher l'explosion de cette technique, parallèlement à l'émergence de la collecte massive des données sur internet. Beaucoup de techniques de profilage ont été développées, sans nécessairement prévoir des garde-fous techniques ou humains. Cette règle est donc peu respectée et sa violation n'a pour l'instant pas donné lieu à sanction.

Parallèlement, le RGPD, et avant lui la directive 95/46/CE, consacre un certain nombre de droits en cas de décision individuelle prise sur le fondement d'un traitement automatisé de données :

1. Le droit d'accès et d'être informé de l'existence d'une prise de décision automatisée (RGDP, art. 13-15h) ;

2. Le droit de ne pas faire l'objet d'un traitement automatisé produisant des effets juridiques ou affectant la personne concernée de manière significative (RGDP, art. 22§1) ;

3. Le droit d'obtenir une intervention humaine de la part du responsable du traitement (RGDP, art. 22§3) ;

4. Le droit d'exprimer son point de vue et de contester la décision (RGDP, art. 22§3) ;

Les données sensibles doivent en principe être exclues des traitements exclusivement automatisés (art. 22§4), sauf en cas de consentement explicite ou pour des raisons d'intérêt public.

Cependant, des exceptions sont aussi prévues (art. 22§2), lorsque la décision :

a) est nécessaire à la conclusion ou à l'exécution d'un contrat entre la personne concernée et un responsable du traitement ;

b) est autorisée par le droit de l'Union ou le droit de L'État membre auquel le responsable du traitement est soumis et qui prévoit également des mesures appropriées pour la sauvegarde des droits et libertés et des intérêts légitimes de la personne concernée ;

c) est fondée sur le consentement explicite de la personne concernée.

Cette série d'exceptions est loin d'être anodine et appauvrit substantiellement la règle. S'agissant des activités économiques du numérique, de nombreux traitements automatisés peuvent en effet se prévaloir d'un fondement contractuel, dès lors que l'utilisation par les internautes des services des sites de e-commerce ou plateformes de mise en relation, telles celles des réseaux sociaux, est de fait considérée comme une acceptation des conditions générales d'utilisation et manifestant l'acceptation de l'offre contractuelle. Par ailleurs, en dehors des activités du numériques, les hypothèses précédemment citées d'accès à un crédit, un logement, à des biens ou services, reposent le plus souvent sur la conclusion d'un contrat.

En outre, le point c) du paragraphe précédent prévoit l'hypothèse d'un consentement explicite de la personne concernée. Si un consentement peut effectivement être assez aisément recueilli en sa forme, on peut toutefois douter au fond de son caractère éclairé, tant l'accessibilité intellectuelle aux procédés de traitement automatisé est douteuse à l'endroit des profanes composant la grande majorité des personnes concernées, spécialement lorsque ce consentement est recueilli en ligne.

Ces dispositions ont été intégrées au droit français avec l'adoption récente de la loi n° 2018-493 du 20 juin 2018 qui vient modifier la loi n° 78-17 dite informatique et libertés du 6 janvier 1978. L'article 21 modifie l'article 10 de la loi du 6 janvier 1978 afin d'étendre les cas dans lesquels, par exception, une décision produisant des effets juridiques à l'égard d'une personne ou l'affectant de manière significative peut être prise sur le seul fondement d'un traitement automatisé de données à caractère personnel. L'article 10 alinéa 1er de la loi n° 78-17 dispose désormais que « Aucune décision de justice impliquant une appréciation sur le comportement d'une personne ne peut avoir pour fondement un traitement automatisé de données à caractère personnel destiné à évaluer certains aspects de la personnalité de cette personne ».

L'alinéa 2 ajoute que « Aucune décision produisant des effets juridiques à l'égard d'une personne ou l'affectant de manière significative ne peut être prise sur le seul fondement d'un traitement automatisé de données à caractère personnel, y compris le profilage ». À ce principe, deux exceptions sont prévues.

La première se réfère aux exceptions du RGPD, c'est-à-dire « les cas mentionnés aux a et c du 2 de l'article 22 précité, sous les réserves mentionnées au 3 de ce même article et à condition que les règles définissant le traitement ainsi que les principales caractéristiques de sa mise en œuvre soient communiquées, à l'exception des secrets protégés par la loi, par le responsable de traitement à l'intéressé s'il en fait la demande ». Outre les garanties prévues par le texte européen à l'article 22§3 (droit d'obtenir une intervention humaine de la part du responsable du traitement, droit d'exprimer son point de vue et de contester la décision), le législateur français a ajouté l'obligation de communiquer les règles définissant le traitement, ainsi que les principales caractéristiques de sa mise en œuvre à la demande de la personne concernée. Cette garantie n'a plus cours si ces règles font l'objet de secrets protégés par la loi. Cette réserve vient ici aussi substantiellement affaiblir le principe, alors même qu'une communication des règles préservant le respect des secrets pourrait aisément s'envisager.

Quant à la deuxième exception prévue à l'article 10 al. 2 de la loi n° 78-17 modifiée, elle s'appuie sur le point b) de l'article 22§2 du RGPD, selon lequel chaque État membre peut prévoir librement des exceptions, dès lors qu'elles sont légalement prévues et respectent certaines garanties. Le législateur français a posé une exception pour les décisions administratives individuelles, à condition que le traitement ne porte pas sur des données sensibles, que des recours administratifs sont possibles et qu'une information est délivrée sur l'usage de l'algorithme. Cette exception ici précisée était déjà consacrée à l'article 4 de la loi n° 2016-1321 pour une république numérique du 7 octobre 2016, codifiée à l'article L. 311-3-1 du CRPA, selon lequel une décision administrative individuelle prise sur le fondement d'un traitement algorithmique doit comporter une mention explicite en informant l'intéressé. L'article 1er du décret n° 2017-330 du 14 mars 2017, codifiée à l'article R. 311-3-1-1 CRPA, précise que la mention explicite doit indiquer la finalité poursuivie par le traitement algorithmique. Elle rappelle le droit d'obtenir la communication des règles définissant ce traitement et des principales caractéristiques de sa mise en œuvre, ainsi que les modalités d'exercice de ce droit à communication et de saisine, le cas échéant, de la commission d'accès aux documents administratifs.

La loi n° 2018-493 du 20 juin 2018 est venue préciser que la mention explicite précitée est exigée à peine de nullité. La sanction de la violation de cette obligation d'information est donc explicitement prévue.

Depuis l'adoption de la loi pour une république numérique le 7 octobre 2016, l'article L. 311-3-1 prévoit par ailleurs que « les règles définissant ce traitement ainsi que les principales caractéristiques de sa mise en œuvre sont communiquées par l'administration à l'intéressé s'il en fait la demande ». Le décret n° 2017-330, codifié à l'article R. 311-3-1-2, précise les informations à fournir sous une forme intelligible et sous réserve de ne pas porter atteinte à des secrets protégés par la loi : 1° Le degré et le mode de contribution du traitement algorithmique à la prise de décision ; 2° Les données traitées et leurs sources ; 3° Les paramètres de traitement et, le cas échéant, leur pondération, appliqués à la situation de l'intéressé ; 4° Les opérations effectuées par le traitement. On constate qu'est maintenue la dérogation en cas de secrets protégés par la loi.

La loi n° 2018-493 va plus loin quant à l'utilisation d'un système de traitement automatisé pour la prise de décision administrative et prévoit désormais une obligation d'explication. Elle dispose ainsi que « le responsable de traitement s'assure de la maîtrise du traitement algorithmique et de ses évolutions afin de pouvoir expliquer, en détail et sous une forme intelligible, à la personne concernée la manière dont le traitement a été mis en œuvre à son égard ». Un fameux « droit à explication » est explicitement consacré par la loi française, alors que le RGPD n'y fait clairement référence que dans le considérant 71. Les articles 13 à 15 se contentent de prévoir un droit d'information et d'accès sur l'utilisation d'un dispositif automatisé et la « logique sous-jacente », ce qui constitue une approche très générale, déconnectée des situations individuelles des personnes concernées.

Par dérogation à cette exception en faveur de l'administration, aucune décision par laquelle l'administration se prononce sur un recours administratif ne peut être prise sur le seul fondement d'un traitement automatisé de données à caractère personnel.

La loi n° 2018-493 a fait l'objet d'une décision du Conseil constitutionnel n° 2018-765 DC le 12 juin 2018, notamment sur les aspects concernant les décisions individuelles automatisées prises par l'administration (points 66 et suivants). Le Conseil constitutionnel estime que les dispositions de la loi se bornent à autoriser l'administration à procéder à l'appréciation individuelle de la situation de l'administré, par le seul truchement d'un algorithme, en fonction des règles et critères définis à l'avance par le responsable du traitement. Elles n'ont ni pour objet ni pour effet d'autoriser l'administration à adopter des décisions sans base légale, ni à appliquer d'autres règles que celles du droit en vigueur. Il n'en résulte dès lors aucun abandon de compétence du pouvoir réglementaire (pt 69).

En deuxième lieu, le seul recours à un algorithme pour fonder une décision administrative individuelle est subordonné au respect de plusieurs conditions. D'une part, la décision administrative individuelle doit mentionner explicitement qu'elle a été adoptée sur le fondement d'un algorithme et les principales caractéristiques de mise en œuvre de ce dernier doivent être communiquées à la personne intéressée, à sa demande. Il en résulte que, lorsque les principes de fonctionnement d'un algorithme ne peuvent être communiqués sans porter at-teinte à l'un des secrets protégés par la loi, aucune décision individuelle ne peut être prise sur le fondement exclusif de cet algorithme (pt 70).

En outre, la décision administrative individuelle doit pouvoir faire l'objet de recours administratifs. L'administration sollicitée à l'occasion de ces recours est alors tenue de se prononcer sans pouvoir se fonder exclusivement sur l'algorithme. La décision administrative est, en cas de recours contentieux, placée sous le contrôle du juge, qui est susceptible d'exiger de l'administration la communication des caractéristiques de l'algorithme. Enfin, le recours exclusif à un algorithme est exclu si ce traitement porte sur l'une des données sensibles mentionnées au paragraphe I de l'article 8 de la loi du 6 janvier 1978, c'est-à-dire des données à caractère personnel « qui révèlent la prétendue origine raciale ou l'origine ethnique, les opinions politiques, les convictions religieuses ou philosophiques ou l'appartenance syndicale d'une personne physique », des données génétiques, des données biométriques, des données de santé ou des données relatives à la vie ou l'orientation sexuelle d'une personne physique (pt 70).

En dernier lieu, le responsable du traitement doit s'assurer de la maîtrise du traitement algorithmique et de ses évolutions afin de pouvoir expliquer, en détail et sous une forme intelligible, à la personne concernée la manière dont le traitement a été mis en œuvre à son égard. Il en résulte que ne peuvent être utilisés, comme fondement exclusif d'une décision administrative individuelle, des algorithmes susceptibles de réviser eux-mêmes les règles qu'ils appliquent, sans le contrôle et la validation du responsable du traitement (pt 71).

Le Conseil constitutionnel estime que le législateur a défini des garanties appropriées pour la sauvegarde des droits et libertés des personnes soumises aux décisions administratives individuelles prises sur le fondement exclusif d'un algorithme. Le 2° de l'article 10 de la loi du 6 janvier 1978 est conforme à la Constitution (pt 72).

Différentes situations peuvent être schématiquement considérées pour l'application de ces règles. Dans le cas d'un algorithme procédural de type *ParcoursSup*, les règles de fonctionnement doivent être clairement explicitées ; le Ministère concerné s'y est préparé à la suite des difficultés rencontrées par le prédécesseur APB. En effet, le code de l'algorithme *Parcoursup* est certes rendu public mais, source d'un débat ou controverse car les règles de délibérations locales à un établissement peuvent rester confidentielles rendant finalement opaque et éventuellement discriminatoire le processus.

Enfin, la loi n° 2018-493 prévoit que, s'agissant plus particulièrement des décisions prises en matière éducative dans le cadre de *ParcoursSup*, « le comité éthique et scientifique mentionné à l'article L.612-3 du code de l'éducation remet chaque année, à l'issue de la procédure nationale de préinscription et avant le 1$^{er}$ décembre, un rapport au Parlement portant sur le déroulement de cette procédure et sur les modalités d'examen des candidatures par les établissements d'enseignement supérieur. Le comité peut formuler à cette occasion toute proposition afin d'améliorer la transparence de cette procédure ».

### 4.2 Quelle transparence ?

Si les dispositions du RGPD ont pour objectif de renforcer les droits des personnes concernées, des lacunes sont constatables, liées, d'une part, aux exceptions et, d'autre part, au fait que l'énoncé de ces droits n'obligent en aucun cas à une forme de transparence. La seule référence à un « droit à explication » prévoit que la personne concernée a le droit d'obtenir du responsable de traitement des informations sur l'existence d'une prise de décision automatisée, y compris un profilage, visée à l'article 22, paragraphes 1 et 4, mais aussi « au moins en pareils cas, des informations utiles concernant la logique sous-jacente, ainsi que l'importance et les conséquences prévues de ce traitement pour la personne concernée ». On peut donc dire que le règlement général sur la protection des données ne concerne pas directement ni véritablement indirectement le principe de transparence algorithmique.

La loi pour une république numérique a essentiellement pour objet d'imposer une obligation d'information (loyauté) sur les modalités de référencement des algorithmes, laquelle s'ajoute aux autres obligations d'information du code de la consommation. Surtout, cette obligation est utilement complétée par les dispositions préexistantes dans le code de consommation relatives aux pratiques commerciales trompeuses dont les énoncés sont suffisamment larges pour viser et sanctionner les comportements déviants qui pourraient être fondés sur des traitements algorithmiques déloyaux ou faussés ; le droit à l'explication ne concerne explicitement que l'administration.

En revanche, la loi n° 2018-493 du 20 juin 2018 pose davantage d'exigences de transparence et explication. Il est encore tôt pour savoir quelle effectivité ces mesures auront à l'avenir sur la transparence algorithmique et l'IA mais le législateur français se montre particulièrement ambitieux, en comparaison du législateur européen et des autres États membres de l'UE.

### 4.3 Quelle explication ?

Dans la suite, nous dirons qu'une décision algorithmique est *interprétable* s'il est possible d'en rendre compte explicitement à partir de données et caractéristiques connues de la situation. Autrement dit, s'il est possible de mettre en relation les valeurs prises par certaines variables (les caractéristiques) et leurs conséquences sur la prévision, par exemple d'un score, et ainsi sur la décision. En revanche, une décision algorithmique est dite *explicable* s'il est seulement possible d'identifier les caractéristiques ou variables qui participent le plus à la décision, voire même d'en quantifier l'importance.

Dans le cas d'un algorithme opaque, il est impossible de mettre simplement en relation des valeurs ou des caractéristiques avec le résultat de la décision, notamment en cas de modèle non linéaire ou avec de nombreuses interactions. Telle valeur élevée d'une variable peut conduire à une décision dans un sens ou dans un autre selon la valeur prise par une autre variable non identifiable, voire même une combinaison complexe d'autres variables. Un modèle opaque qui ne permet pas de s'expliquer facilement, par exemple face un candidat à l'embauche, aboutit à une forme de déresponsabilisation du décideur lui permettant de se cacher derrière l'algorithme. Ce n'est plus la faute de l'ordinateur mais celle de l'algorithme.

Chaque acteur, public ou privé, chaque domaine, santé, justice, emploi, banque, assurance, police… nécessite une lecture spécifique de ce que peut être une forme de transparence algorithmique en lien avec le droit à l'explication. Un comportement éthique *a minima* est indispensable à l'acceptabilité de ces technologies mais, dans ce cadre, la formulation d'une explication dépend de bien des facteurs. Expliquer un diagnostic automatique et les risques relatifs encourus lors d'une intervention chirurgicale, motiver une mesure de détention dépendant d'une estimation algorithmique d'un risque de récidive, justifier le refus d'un prêt sur la base d'un score… nécessitent une double compétence : compétence métier et connaissance des limites, propriétés, de l'algorithme qui a conduit à la décision.

### 4.4 Aides à l'explication

Que peut faire un individu, qu'il soit responsable d'une décision ou simple citoyen, client, concerné par celle-ci, confronté à un ensemble de quelques centaines d'arbre de décision, d'un réseau de neurones définis par des milliers voire des millions de paramètres appris sur un gros volume de données ?

Dans beaucoup de domaines d'application et notamment en médecine, un modèle opaque qui ne permet pas de s'expliquer facilement, par exemple face un patient, et qui aboutirait à une forme de déresponsabilisation du décideur, ne serait que difficilement acceptable, à moins d'apporter une qualité de prévision nettement supérieure dans la recherche d'un meilleur *compromis entre qualité et explicabilité*. En d'autre terme faut-il privilégier un modèle élémentaire interprétable mais éventuellement moins précis à un modèle complexe, faisant intervenir de très nombreux paramètres et possédant de meilleures qualités prédictives mais opaque à toute interprétation.

Une approche graduée pourrait s'envisager, suivant la priorité donnée à l'explication ou la qualité de la prévision, à supposer qu'un algorithme plus opaque permette *a contrario* de meilleurs résultats. La réponse serait alors potentiellement différente selon les activités concernées car il n'est pas pertinent de traiter de la même façon un algorithme utilisé en médecine ou en techniques de commercialisation. Cela conduirait alors à encourager une réglementation sectorielle. En tout état de cause, il paraît indispensable de pouvoir faire un choix social sur ce qui est préférable dans une balance d'intérêts circonstanciés entre la qualité de l'explication et la qualité de la prévision, au moins dans les hypothèses où les caractéristiques des algorithmes sont réductibles à ces deux principales qualités.

Notons par ailleurs que le « droit à explication » peut faire l'objet de deux approches différentes (Watcher et a. 2017 p5) :

- le droit d'avoir une explication sur le fonctionnement général du système mettant en œuvre des décisions algorithmiques ;
- le droit d'avoir une explication sur une décision spécifique.

Au demeurant, l'explication peut être *ex ante* ou *ex post* (Watcher et al. 2017 p6). S'il s'agit de donner une explication spécifique sur une décision individuelle, l'explication ne pourra être donnée que *ex post*, alors que si elle porte sur le fonctionnement général, elle pourra l'être *ex ante* ou *ex post*.

Dans le cas d'un algorithme d'apprentissage interprétable, les coefficients d'un modèle linéaire ou logistique peuvent et doivent être explicités pour l'individu concerné de même que la séquence de règles définissant un arbre de décision. Le dernier cas d'un algo-rithme opaque ou seulement explicable semble difficilement concerné ou pris en compte par la loi.

Compte tenu de l'importance des enjeux sur l'explicabilité (ouvrir les boîtes noires de l'IA), la recherche est très active en ce domaine. Citons quelques exemples illustratifs selon qu'une compréhension globale d'un algorithme complexe ou qu'une explication individuelle est visée. Dans le premier cas, des aides visent à identifier les facteurs, variables (*features*) les plus importants c'est-à-dire ceux participant généralement le plus à une décision. Indépendamment de toute question éthique, cela est fondamental pour analyser la fiabilité ou la robustesse de décisions et identifier les possibles artefacts généralement dus aux insuffisances de la base d'apprentissage. Différentes stratégies sont proposées ; ainsi, pour les algorithmes d'agrégation d'arbres (*random forest, gradient boosting*) il est d'usage de chercher les variables dont une permutation circulaire des valeurs (*mean decrease accuracy*) dégrade le plus une estimation de la qualité de prévision. Une méthode plus générale consiste à approcher localement les décisions obtenues par un algorithme non interprétables au moyen d'une règle de décision interprétable de type régression. En effet pour une régression ou un modèle de régression logistique le rôle joué par chaque variable est clairement exprimé au moyen d'une combinaison linéaire. Chaque coefficient correspond au poids de chaque variable dans la prédiction et permet ainsi de déterminer non seulement l'importance de chaque variable mais également si sa contribution est positive ou négative dans le résultat final. La règle de décision pouvant être beaucoup plus complexe qu'une règle linéaire, cette approximation n'a pas de sens globalement mais seulement localement. Cette méthodologie est développée dans le package LIME (2016). Une idée similaire consiste à tester l'algorithme sur un algorithme présentant un biais pour chaque variable. Ainsi si l'on parvient à créer un échantillon de test de loi presque similaire mais pour lequel une variable présente une déviation de sa moyenne (positive ou négative), on peut étudier l'évolution de la règle de décision de manière globale puisque l'on considère les lois de l'échantillon des sorties de l'algorithme avec pour entrées l'échantillon de test modifié. Cette méthodologie permet de répondre à la question : comment peut-on influer en moyenne sur une décision prise par un algorithme en augmentant ou en diminuant certaines de ses caractéristiques. Ce travail est détaillé dans Bachoc et al. (2018).

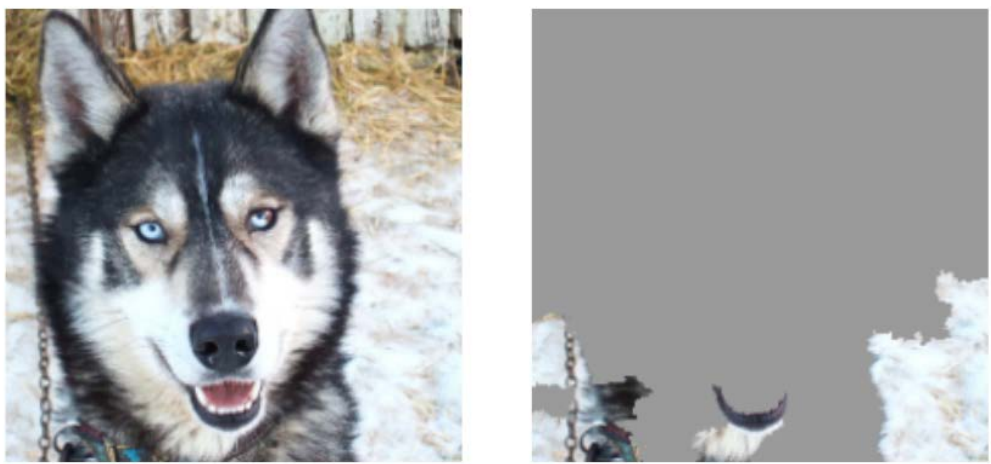

*Figure 2. Un huski (à gauche) est confondu avec un loup car les pixels (à droite) caractéristiques de ces derniers sont ceux du fond de neige. Cet artefact est dû à une base d'apprentissage pas suffisamment représentative.*

Cette approche est encore valide pour les algorithmes même très complexes, apprentissage profond (*deep learning*), utilisés en reconnaissance d'images. Un exemple souvent repris (Tulio Ribeiro et al. 2016) met en évidence une coquille de la base d'apprentissage. Comme des loups ont systématiquement été photographiés sur un fond de neige, c'est le fond qui permet de les identifier le plus sûrement. Un husky lui aussi photographié devant de la neige est alors confondu avec un loup (figure 2).

Le problème se pose différemment pour l'explication d'une décision individuelle. Il s'agit encore d'identifier les ou plutôt la variable la plus représentative au sens qu'une modification de celle-ci permettrait de faire basculer la décision. Par exemple, quel revenu serait nécessaire à l'obtention d'un prêt, ou quelle variable comportementale impacte le plus favorablement l'évaluation d'un score de récidive. Cet objectif pourrait être atteint par une approximation locale interprétable de la règle de l'algorithme complexe ; approximation par un modèle linéaire simple ou un arbre binaire de décision élémentaire.

En conclusion, le règlement européen et la loi française, principalement axée sur l'administration, laissent un large espace de manœuvre en matière de transparence. Espace vide qui, sans progrès significatif de la recherche fondamentale sur ce sujet, doit être occupé par des actions éthiques au risque de provoquer un rejet massif des technologies de l'IA. Ainsi, le partenariat (Partnership on IA) entre les principaux acteurs (Google, Facebook, Microsoft, IBM, Apple…) pour une IA au service de l'homme est très sensible à ce besoin d'interprétation. Un article de leur charte précise :

*7. We believe that it is important for the operation of AI systems to be understandable and interpretable by people, for purposes of explaining the technology.*

## 5 Qualité de décision et erreur de prévision

### 5.1 Mesurer une qualité de décision

En apprentissage statistique, la *justesse* de la décision dépend de la qualité d'une prévision et donc de la qualité d'un modèle ou d'un algorithme. Cette dernière dépend de la *représentativité* ou biais des données initiales, de l'adéquation du modèle au problème posé et à la quantité (variance) de bruit résiduel. Elle est évaluée sur un échantillon test indépendant ou par validation croisée (*Monte Carlo*) mais reste indicative sous la forme d'un *risque probabiliste d'erreur.*

Les méthodes de prévisions sont entraînées sur les données d'apprentissage, c'est donc la *qualité* de celle-ci qui est en premier lieu déterminante ; rappelons le vieil adage : *garbage in, garbage out*. Leur volume peut être un facteur utile de qualité mais seulement si les données sont bien représentatives de l'objectif et pas biaisées. Dans le cas contraire des téraoctets n'y font rien. C'est l'exemple de [Google Flu Trend](#) (2008) qui visait à suivre en temps réel et prédire le déroulement d'une épidémie de grippe à partir du nombre de recherches de certains mots clefs associés et connaissant la localisation (adresse IP) du questionneur. L'outil a été abandonné par Google (2015) car source de lourdes erreurs de prévision. C'était le battage médiatique de la grippe qui était suivi, pas l'épidémie elle-même. Les données ont été reprises avec de meilleurs résultats par une équipe de Boston (Yanga et al, 2015) en estimant un modèle auto-régressif intégrant une chaîne de Markov cachée et corrigée sur la base des tendances des recherches sur Google.

La littérature sur l'apprentissage automatique est excessivement prolifique sur les façons de mesurer et estimer les erreurs. Il faut distinguer en premier lieu l'erreur d'ajustement ou d'apprentissage qui qualifie le bon respect des données et celle à proprement parler de prévision ou de généralisation. Le type de mesure ou fonction perte est adapté au type de la variable cible ou à prévoir ; celle-ci peut être réelle, quantitative (problème de régression) ou discrète, qualitative (classification supervisée). Dans le cas quantitatif, la fonction perte peut être quadratique (erreur quadratique moyenne) ou basée sur une valeur absolue, plus robuste aux valeurs atypiques mais aussi plus complexe à optimiser. Dans le cas qualitatif, il peut s'agir d'un simple taux d'erreur ou d'une mesure d'entropie ou, notamment dans le cas de classes déséquilibrées, de mesure plus complexes. L'erreur d'apprentissage ajoute généralement une pénalisation à la fonction perte afin de contrôler, par la valeur d'un paramètre à optimiser, la complexité du modèle dans l'objectif d'atteindre un meilleur compromis biais *vs.* variance pour échapper au sur-apprentissage. En définitif, une fois l'algorithme entraîné et optimisé sur l'échantillon d'apprentissage, c'est l'estimation de l'erreur de prévision sur un échantillon test indépendant et de taille suffisante qui donne une indication de la qualité d'une décision algorithmique.

### 5.2 Enjeu de la qualité d'une décision algorithmique

Il est notable que l'erreur de prévision impacte le biais ou caractère discriminatoire d'une décision (cf. section 3.5 de l'exemple du score de récidive) et influence le choix d'une méthode ou d'un algorithme dans la recherche d'un meilleur compromis entre précision et interprétabilité. Bien que fondamentale à bien des égards, notamment pour être en mesure de discuter de l'opportunité d'une décision (*e.g.* conséquence d'un diagnostic médical), la loi comme le RGPD n'en font absolument aucune mention. Il semblerait en effet très pertinent qu'une décision algorithmique soit accompagnée d'une évaluation du risque d'erreur encouru comme la loi oblige les instituts de sondage à produire des marges d'incertitude.

Les principaux fournisseurs ou vendeurs d'Intelligence Artificielle (Google, Facebook, IBM, Microsoft…) ont intérêt à mettre en évidence, amplifiés par les médias, les résultats les plus spectaculaires de l'IA (reconnaissance d'images, traduction automatique, compétition de jeu de go…) avec des taux de succès exceptionnels, meilleurs que l'expert humain. Mais ces succès portent principalement sur des prototypes, ou sur des applications sans enjeux risqués. Malheureusement (quoique ?) les taux d'erreurs attachés à la prévision de comportements humains (score de récidive d'un détenu, détection de commentaires injurieux, de fausses nouvelles, d'un comportement à risque, etc.) sont nettement plus, voire tristement, pessimistes.

La loi ne codifie pas une obligation de résultat mais une pratique éthique ou un manuel de bonne pratique de conception impose au concepteur, comme au médecin, une obligation de moyens : tout mettre en œuvre afin d'assurer au citoyen, client, patient… la meilleure décision possible en l'état des connaissances. Par ailleurs, l'évaluation de l'erreur, la répartition de ses causes contribueraient efficacement à la réflexion sur le *partage des responsabilités*. L'obligation de publication ou de notification de la qualité de l'algorithme utilisé serait, comme pour les sondages, un facteur important de responsabilisation de l'utilisateur.

## 6 Conclusion

La prise de conscience progressive de la puissance que peuvent avoir des systèmes de décision automatiques utilisant des techniques d'apprentissage statistique pour exploiter les masses de données désormais omniprésentes dans tous les secteurs d'activité (commerciales, administratives, économiques, industrielles, médicales, etc.) suscite autant d'espoirs que de craintes légitimes. On ne peut compter exclusivement sur la responsabilité des acteurs de ces changements, ni sur la dynamique propre du front de recherche en apprentissage automatique, pour éviter les dérapages, voire la banalisation d'usages abusifs de ces techniques. Les risques portent notamment sur les discriminations, sur l'arbitraire de décisions dont on ne connaît guère la pertinence et dont on ne sait trop qui est responsable, sur les dérives d'un développement purement guidé par les possibilités techniques, et sur les biais, même involontaires, induits par le processus de récolte des données qui conditionne le comportement des algorithmes. Ils portent également sur des points non abordés dans ce court article : confidentialité des données et risques de ré-identification, entrave à la concurrence.

La principale difficulté vient de ce qu'aborder sérieusement ces questions nécessite à la fois de sérieuses compétences techniques, afin de comprendre finement le fonctionnement des algorithmes et de garder un regard critique sur le discours qui les entoure, et une expertise juridique, sociétale ou sociologique, voire politique ou philosophique. La teneur des débats sur le sujet, et l'analyse des textes juridiques même récents, montrent que le défi est grand.

- Une pratique discriminatoire envers une personne est punie par la loi mais il revient à la victime d'en apporter la preuve. Aucun texte ou jurisprudence ne définit, contrairement aux USA, ce que pourrait être une discrimination ou une mesure de discrimination (DIA du rapport Villani) envers un groupe.

- L'obligation de transparence ou d'explicabilité, impose au mieux une intervention humaine pour assumer une décision et n'est contraignante que pour les décisions administra-tives françaises, interdisant ainsi l'usage d'algorithmes auto-apprenants sans contrôle ou validation humaine, comme ce peut être le cas de ventes de publicités en ligne.

- Aucun texte n'oblige à publier ou renseigner la qualité de prévision ou le taux d'erreur associé à l'utilisation d'un algorithme d'apprentissage.

La disruption technologique qui en découle autorise toutes les possibilités de comportements ou de pratiques, éthiques ou pas. Les questions de discrimination sont celles le mieux encadrées par la loi mais aussi celles les plus complexes à appréhender. L'exemple souvent cité de justice prédictive (score de récidive) montre bien que les décisions qui en découlent ne peuvent être que largement statistiquement biaisées donc collectivement discriminatoires, sur certains critères, mais sans pour autant qu'il soit facile, pour une personne, de montrer qu'elle en a été lésée. Cet exemple montre par ailleurs très bien que les données, bases de l'apprentissage des algorithmes, et leurs modes sélectifs de recueil, reflets de nos sociétés, sont la principale source d'erreurs ou de biais.

Cette situation motive en retour la recherche fondamentale pour définir des modèles ou construire des algorithmes répondant à ces critiques. Les investigations en cours consistent à rechercher des meilleurs compromis entre différentes contraintes : explicabilité et qualité de prévision, réduction du biais et confidentialité des données.

Vérifier l'interprétabilité ou l'explicabilité d'un algorithme ou du modèle sous-jacent, contrôler, par exemple sur un échantillon test, ses qualités prédictives et enfin détecter des biais potentiels, collectifs ou individuels sont des tâches complexes. À l'heure actuelle, aucun acteur ne peut à lui seul prétendre pouvoir contrôler la loyauté algorithmique. Une pluralité de contre-pouvoirs est donc nécessaire. Quels sont les acteurs susceptibles de prendre en charge ces contrôles ? Certains sont les régulateurs publics : CNIL, DGCCRF (répression des fraudes), Autorité de la Concurrence, juges (juridictions françaises et CJUE) mais en ont-ils les moyens ? D'autres sont privés : plateformes collaboratives (Data transparency lab, TransAlgo INRIA, Conseil National du Numérique), Médias (ProPublica aux USA), ONG Data (*Bayes Impact*) mais ne sont que balbutiants et difficilement finançables.

Faut-il aller plus loin que les principes énoncés par la loi pour une république numérique et la loi de 1978 modifiée par la loi n° 2018-493 ? À l'heure actuelle, il convient d'abord de laisser le temps à ces lois de s'appliquer pour en mesurer la portée. Cela peut paraître prématuré, alors même que l'efficacité des dispositifs de contrôle est encore incertaine. En outre, comment formuler plus précisément les conditions d'encadrement dans l'utilisation des différentes méthodes algorithmiques alors que le domaine d'application, commerciale, juridique, médical, administratif… en change considérablement l'environnement et donc les conditions d'utilisation ou les termes d'explication ?

Dans ce contexte encore flou, il paraît peu pertinent de s'en remettre une nouvelle fois et dès à présent au législateur, sauf pour éventuellement réclamer, comme pour les sondages, l'obligation d'affichage d'un taux d'erreur. Les recherches académiques émergent seulement depuis 2-3 ans et il convient de prendre un peu de recul avant d'imposer une règle précise à respecter. D'autres normes commencent à apparaître, simples règles éthiques, bonnes pratiques (*soft law*), qui pourraient aider à mieux cerner les conditions d'une loyauté et transpa-rence algorithmique : Cathy O'Neil évoque la nécessité d'un serment d'Hippocrate pour les *data scientists,* idée reprise par de très nombreux groupes ou associations dont celle française *data for good*, tandis qu'un groupe de travail offre de signer un code de conduite sur l'éthique des pratiques des données… les initiatives se sont multipliées et il serait difficile d'en établir une liste exhaustive. Notons que les statisticiens européens ou américains se sont dotés de longue date de chartes ou codes de bonnes pratiques mais ces textes ne peuvent être transposés sans réflexion approfondie comme par exemple sur la notion de *consentement libre et éclairé* lorsque celui-ci est recueilli en ligne.

Enfin, la pratique de l'IA au quotidien ayant besoin de confiance de la part des usagers, confiance qu'il est difficile d'accorder sans contrôle aux fournisseurs et vendeurs de technologies, une autre solution consiste à proposer aux entreprises et principaux acteurs la délivrance d'un label indépendant témoignant, à la suite d'un audit, d'un usage loyal des données (*fair data use*). C'est ce que propose la société ORCAA créée par Cathy O'Neil ou encore la startup Maathics.

## Références